\documentclass[apj,iop]{emulateapj}
\usepackage[utf8]{inputenc}
\usepackage[english]{babel}
\usepackage[T1]{fontenc}
\usepackage{ mathrsfs }
\usepackage{amsmath}
\usepackage{gensymb}
\usepackage{natbib}

\begin{document}
\title{Source-dependent properties \\
of two slow solar wind states}


\author{Léa Griton}
\email{lea.griton@irap.omp.eu}
\affiliation{IRAP, Universit\'e Toulouse III - Paul Sabatier,
CNRS, CNES, Toulouse, France}

\author{Alexis P. Rouillard}
\affiliation{IRAP, Universit\'e Toulouse III - Paul Sabatier,
CNRS, CNES, Toulouse, France}

\author{Nicolas Poirier}
\affiliation{IRAP, Universit\'e Toulouse III - Paul Sabatier,
CNRS, CNES, Toulouse, France}

\author{Karine Issautier}
\affiliation{LESIA, Observatoire de Paris, Universit\'e PSL, CNRS, Sorbonne Universit\'e, Universit\'e de Paris, 5 place Jules Janssen, 92195 Meudon, France}

\author{Michel Moncuquet}
\affiliation{LESIA, Observatoire de Paris, Universit\'e PSL, CNRS, Sorbonne Universit\'e, Universit\'e de Paris, 5 place Jules Janssen, 92195 Meudon, France}

\author{Rui F. Pinto}
\affiliation{IRAP, Universit\'e Toulouse III - Paul Sabatier,
CNRS, CNES, Toulouse, France}
\affiliation{LDE3, CEA Saclay, Université Paris-Saclay, Gif-sur-Yvette, France}

\begin{abstract}
Two states of the slow solar wind are identified from in-situ measurements by Parker Solar Probe (PSP) inside 50 solar radii from the Sun. At such distances the wind measured at PSP has not yet undergone significant transformation related to the expansion and propagation of the wind. We focus in this study on the properties of the quiet solar wind with no magnetic switchbacks. The two states differ by their plasma beta, flux and magnetic pressure. PSP's magnetic connectivity established with Potential Field Source Surface (PFSS) reconstructions, tested against extreme ultraviolet (EUV) and white-light imaging, reveals the two states correspond to a transition from a streamer to an equatorial coronal hole. The expansion factors of magnetic field lines in the streamer are 20 times greater than those rooted near the center of the coronal hole. The very different expansion rates of the magnetic field result in different magnetic pressures measured by PSP in the two plasma states. Solar wind simulations run along these differing flux tubes reproduce the slower and denser wind measured in the streamer and the more tenuous wind measured in the coronal hole. Plasma heating is more intense at the base of the streamer field lines rooted near the boundary of the equatorial hole than those rooted closer to the center of the hole. This results in a higher wind flux driven inside the streamer than deeper inside the equatorial hole.

\end{abstract}

\keywords{Solar corona --- 
Solar wind --- Magnetohydrodynamical simulations
}

\date{Accepted for publication in The Astrophysical Journal, on Feb 4, 2021}

\section{Introduction}
\label{section:introduction}

The ambient solar wind is made of a variety of streams and structures whose properties cover a wide range of values depending on several aspects: origin of the wind at the Sun, but also acceleration processes, turbulence, and stream-stream interactions (see review by \cite{cranmer_origins_2017}). It is therefore challenging to relate solar wind bulk properties measured near 1 AU (e.g. \cite{iii_statistical_2018}) to clearly defined coronal sources. Composition and suprathermal electron measurements have remained for a long time the most relevant wind measurements to probe source-dependent effects \citep{geiss_origin_1995}.
Parker Solar Probe (PSP) is measuring solar wind plasma and magnetic field with both remote and in situ instruments, closer to the Sun than any probe before since its first perihelion on November 5, 2018. During encounter 2 (E2), whose perihelion took place on April 5, 2019, PSP measured essentially slow solar wind (with a speed slower than $500~\mathrm{km.s^{-1}}$).

The origin of slow solar wind is still debated. Sources of slow solar wind could be the boundaries of coronal holes, along magnetic flux tubes expanding super-radially \citep{wang_solar_1990,strachan_empirical_2002,antonucci_slow_2005, stakhiv_origin_2015, arge_improved_2003}. The slow solar wind appears to have different components, which have been related to different properties at the very source of the wind, for example Alfvénicity, which is thought to be higher in the slow solar wind emerging from coronal holes than in the slow solar wind originating closer to the edge of streamers \citep{damicis_origin_2015, ko_boundary_2018, perrone_highly_2020}.  Slow solar wind plasma could also originate from multiple reconnection processes, known as "interchange reconnection"  \citep{fisk_coronal_1999,wang_dynamical_2000,antiochos_model_2011} occurring at the edge of streamers or pseudo-streamers \citep{crooker_comparison_2014}, that could produce a more time-variable solar wind, along with other phenomena which are detailed in the literature (see for example section 3.2 of \cite{cranmer_origins_2017} or\cite{xu_new_2015}).

During E2, PSP crossed a boundary between two different plasma states on April 3, 2019 at 07:55 UT. This boundary crossing was studied observationally by comparing PSP in situ data with white-light images from Stereo-A and SOHO \citep{rouillard_relating_2020}. For a few days before April 3, 2019,  07:55 UT, PSP was sampling a dense, slow and highly variable solar wind (state 1). However, after April 3, 2019,  07:55 UT, and until April 5, 2019, 23:59 UT, the wind became suddenly less dense, less variable, and increasingly faster (state 2). PSP's first encounter with the Sun revealed that the slow solar wind close to the Sun is highly variable consisting of a profusion of magnetic field 'switchbacks' \citep{bale_highly_2019} and plasma jets \citep{kasper_alfvenic_2019}. In this paper, we explore in more detail the magnetic field and plasma properties of those two different plasma states during E2. We focus our analysis on a subset of time intervals of very quiet solar wind that are not perturbed by switchbacks. We found those intervals in both plasma states. Our aims were twofold. First we wanted to investigate whether the properties of the calm solar wind can be interpreted by wind models that do not require transient processes to produce a slow wind. Second, we searched for the origin of the sharp transition in wind properties measured on April 3, 2019, 07:55 UT.

This is investigated through a detailed analysis of the data taken by the Fields Experiment (FIELDS) and the Solar Wind Electrons Alphas and Protons (SWEAP) Investigation. To establish a link between the \textit{in situ} measurements and the structure of the magnetic field at the surface of the Sun, we exploited Potential Field Source Surface (PFSS) reconstructions of the coronal magnetic field, tested against white light images from SOHO-LASCO and Extreme UV observations from SDO-AIA (193\AA). 

The paper is structured as follows. In section \ref{section:m&m}, we describe the \textit{in situ} data measured by Parker Solar Probe (PSP) that are exploited in this study. In section \ref{section:results}, we present the results of the data analysis. We detail the method employed to study the connectivity of the spacecraft on the given time interval and present the set-up of numerical simulations in section \ref{section:m&m-connectivity}. In section \ref{section:discussion}, we relate the different properties measured in the two slow winds by combining the magnetic field reconstructions with solar wind modelling. A summary of the key findings is given in the conclusion, in section \ref{section:conclusion}. \\

\section{Methods for data analysis}
\label{section:m&m}
\subsection{Instruments and data processing}
\label{section:m&m-data}

All data exploited in this study were obtained by the Parker Solar Probe (PSP) \citep{fox_solar_2016}. Data include magnetic field components in the Radial Tangential Normal (R,T,N) coordinates with respect to the spacecraft (provided by the FIELDS instrument, \cite{bale_fields_2016}). Electron density and core temperature are provided by the Quasi-Thermal-Noise spectroscopy \citep{meyervernet_quasi-thermal_2017} applied to FIELDS data \citep{moncuquet_first_2020}. Proton speeds were measured by the Solar Probe Cup (SPC), which is part of the Solar Wind Electrons Alphas and Protons (SWEAP) instrument \citep{kasper_solar_2016}.\\

\begin{figure*}[!h]
\includegraphics[width=0.95\linewidth]{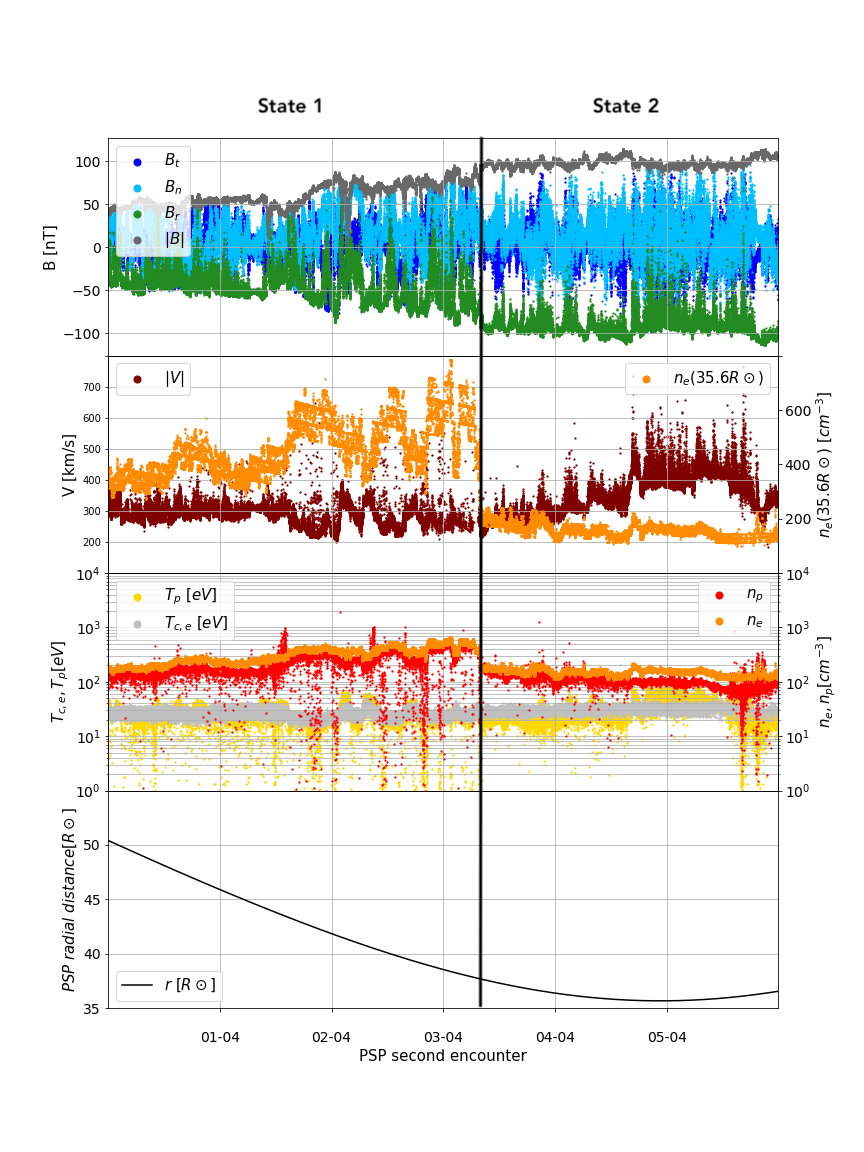}
\caption{All data sets used within this study, from March 31, 2019, 00:00 UT to April 5, 2019, 23:59 UT. Top panel: the original magnetic field components measured by FIELDS in (R,T,N) coordinates (not scaled), along with the total magnetic field intensity (grey line), all in nanoTesla. Second panel: the proton speed measured by SWEAP/SPC (brown) (not scaled) along with the electron density (FIELDS/RFS(QTN)) , scaled to the perihelion distance at 36.5 $R\odot$. Third panel: original proton and electron densities in particles per $cm^{-3}$, along with original electron core temperature and proton temperature (not scaled and deduced from proton thermal speed). Bottom panel: the heliocentric radial distance of the spacecraft as a function of time.}		
\label{Fig2-1}
\end{figure*}

Data sets cover the period from March 31, 2019, 00:00 UT to April 5, 2019, 23:59 UT (during Carrington rotation 2215), during PSP's second encounter with the Sun. They are shown on Figure \ref{Fig2-1}. During this time interval, the spacecraft travelled from 50.4 $R\odot$ to 35.6 $R\odot$, with the spacecraft perihelion on April 5, 10:40 UT at 35.6 $R\odot$. We used the time provided by the plasma QTN spectrum measured by RFS/FIELDS (electron densities and temperatures \citep{moncuquet_first_2020} as a reference (most of the time, QTN data are provided every 7 seconds). We interpolated the magnetic field (FIELDS) and proton (SCP) data, all averaged on 1 second intervals, to the QTN time using the closest value approach. 

\begin{figure*}[!h]
	\includegraphics[width=\linewidth]{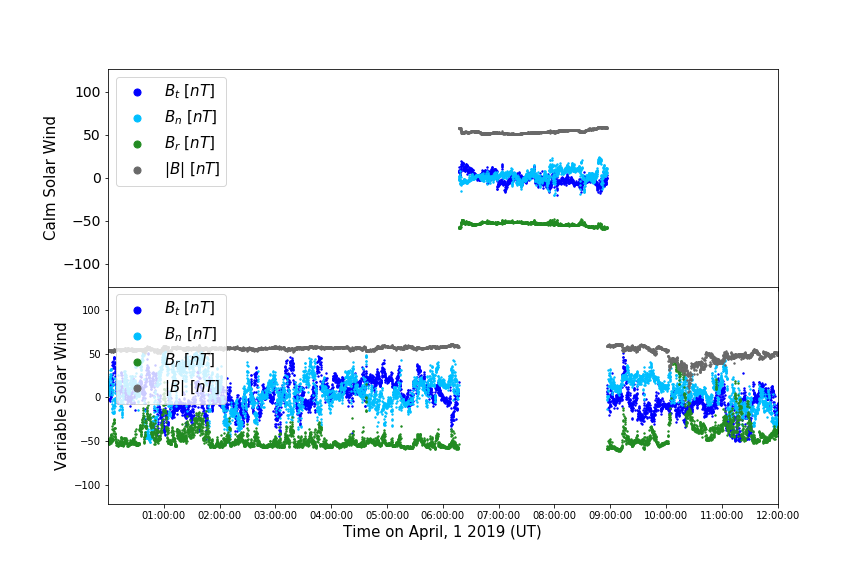}
	\caption{Magnetic field data on April 1, 2019 from 00:00 UT to 12:00 UT, with a calm solar wind period shown in the top panel, in between two periods of variable solar wind that include magnetic switchbacks. }
	\label{Fig0-CSW_intervals}
\end{figure*}

\begin{figure*}[!t]
		\includegraphics[width=\linewidth]{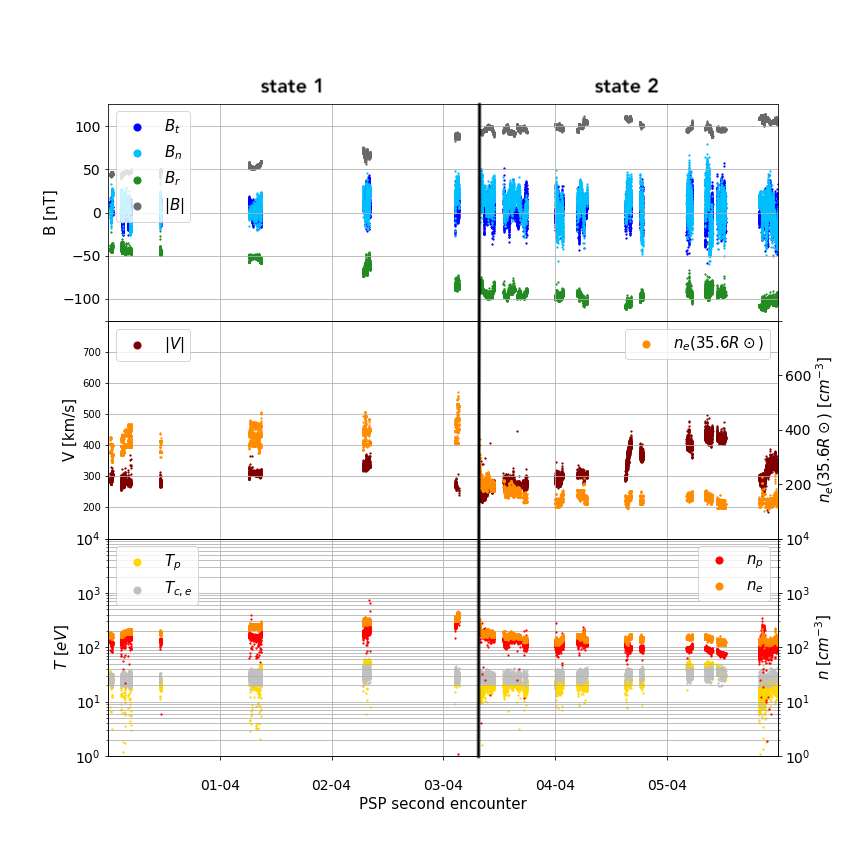}
		\caption{Same as the previous figure but with only the data points from the "calm" solar wind intervals. Those intervals were manually checked not to contain any magnetic switchbacks.
		}
		\label{Fig2-2}
	\end{figure*}

We manually selected time intervals with no magnetic field switchbacks, based on the selection of long-enough time intervals (i.e. more than 30 minutes) without strong variation of the radial component of the magnetic field. These intervals were determined by visual inspection of the magnetic field data. An example of such a 'calm' interval that occurred between two intervals of variable wind perturbed by magnetic switchbacks is shown in Figure \ref{Fig0-CSW_intervals}. The periods of less variable solar wind are referred to as the "calm solar wind intervals". The calm solar wind intervals are shown in Figure \ref{Fig2-2}.

\subsection{Removing the radial evolution}
\label{section:m&m-radialevolution}

The aim of this study being to differentiate between different sources of the slow wind, we corrected the data to remove the radial variation of the different parameters. At first order and at constant speed, the density of the spherically expanding solar wind is expected to decrease as the inverse of the square of the radial distance from the Sun. We thus provide densities at a reference perihelion distance of 35.6 $R\odot$ by multiplying values by $(r/35.6)^2$, where $r$ is the radial distance of PSP from the Sun given in solar radii ($R\odot$). The magnitude of the magnetic field is also related to its estimated value at 35.6 $R\odot$, taking into account the decrease in $1/r^2$ due to spherical expansion. The radial decrease of the core electron temperatures is directly taken from \cite{moncuquet_first_2020}: $T_{c,e} \propto (r/R\odot)^{-0.74}$.

\section{Identification and qualification of two distinct plasma states}
\label{section:results}

\subsection{A change in the dominant pressure in the two plasma states}

\begin{figure*}[!h]
	\includegraphics[width=\linewidth]{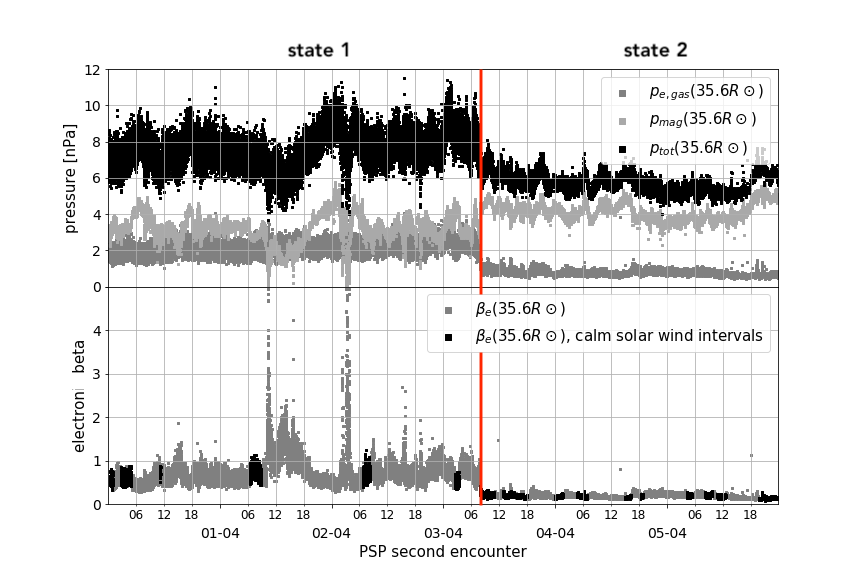}
	\caption{Top panel: total pressure $p_{tot}=2n_ek_BT_{c,e}+p_{mag}$ in black, along with the electron gas pressure  $p_{e}=n_ek_BT_{c,e}$  in dark grey and magnetic pressure $p_{mag}=|\textbf{B}|^2/2\mu_0$, computed with all quantities scaled to the perihelion distance of $35.6 R\odot$ (see sub-section\ref{section:m&m-radialevolution}), all in nPa. Bottom panel: electron $\beta_e=n_ek_BT_{c,e}/p_{mag}$, in black for calm solar wind intervals, and in grey otherwise. Here quantities are also scaled to $35.6 R\odot$.}
	\label{Fig3-beta1}
\end{figure*}

The plasma $\beta$ is defined as the ratio of the plasma gas pressure to the magnetic pressure. Thus $\beta$ describes the dominant pressure state in the plasma: $\beta > 1  $ means that the gas dominates the magnetic pressure, and vice versa. On April 3, 2019, at 07:55 UT, Parker Solar Probe encounters a boundary between two different plasma states, as  seen in Figure \ref{Fig3-beta1}. Only the electron $\beta$ is considered here, defined as $\beta_e=n_ek_BT_{c,e}/p_{mag}$, with $n_e$ the electron density, $k_B$ the Boltzmann constant, $T_{c,e}$ the core electron temperature and $p_{mag}$ the magnetic pressure: $p_{mag}=|\textbf{B}|^2/2\mu_0$, where $\textbf{B}$ is the magnetic field and $\mu_0$ the vacuum permeability. The total $\beta$, including the proton, alphas and heavier ions contributions to the gas pressure, would be approximately twice higher than the electron $\beta_e$ (if the fraction of the alpha particles and heavy ions compared to protons can be neglected). We see that the $\beta_e$ mean value, computed on the calm solar wind intervals, is $\simeq 0.7$ in state 1 (standard deviation: 0.26 in state 1), which is close or superior to $0.5$. Then the $\beta_e$ mean value computed on the calm solar wind intervals suddenly drops to $\simeq 0.2$ (standard deviation: 0.06 in state 2) , which is much lower than $0.5$. The total pressure is estimated as $p_{tot}=2n_ek_BT_{c,e}+p_{mag}$ and is displayed in the top panel of Figure \ref{Fig3-beta1}. The total pressure is less variable in state 2 (standard deviation is 0.95 in state 1 and 0.53 in state 2), and decreases slightly, by a factor of $0.25$: in state 1, the median value of $p_{tot}$ is $\simeq$7.6 nPa, and it falls to 5.7 nPa in state 2. One may note here that  $p_{tot,2}=n_ek_BT_{c,e}+n_pk_BT_{p}+p_{mag}$, using SWEAP/SCP proton density and temperature, follows the same trend as $p_{tot}$ but shows greater dispersion, due to calibration issues. However, a clear boundary is less visible in the total pressure, contrary to what is observed for the electron  $\beta_e$ (bottom panel of Figure \ref{Fig3-beta1}), suggesting that the states identified here are roughly in pressure equilibrium at the boundary. When looking at the electron gas pressure and magnetic pressure separately (top panel of Figure \ref{Fig3-beta1}), the transition is clearly visible for each pressure, and is especially sharp for the electron gas pressure.\\

\begin{figure}[!h]
	\includegraphics[width=0.9\linewidth]{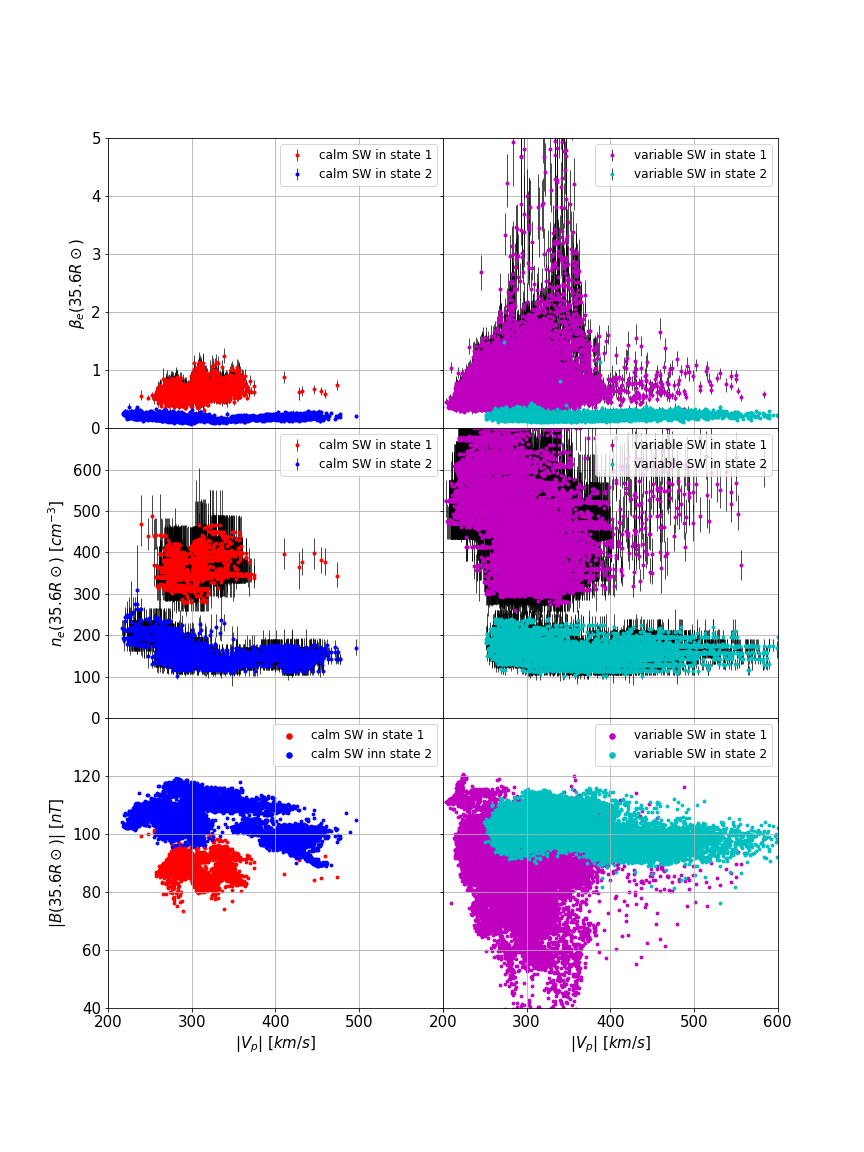}
	\caption{Left column: data for calm solar wind intervals only, in red for state 1 and in blue for state 2. Right column: rest of the data-set, in magenta for state 1 and in cyan for state 2. Top panels: electron $\beta_e$ scaled to $35.6\ R_\odot$ (see sub-section\ref{section:m&m-radialevolution}), against proton speed in km/s.  Middle panels: electron density [$cm^{-3}$] scaled to $35.6\ R_\odot$, against proton speed in km/s. Error bars provided by the QTN method are plotted as thin black lines behind each point and are used to compute the error on $\beta_e$. Bottom panels: magnetic field intensity [nT] scaled to $35.6\ R_\odot$, against proton speed in km/s.}
	\label{Fig3-beta2}
\end{figure}

When plotting the electron $\beta_e$ against the proton bulk speed (provided by SWEAP/SCP), which is used as a proxy for plasma speed, a clear difference appears from state 1 to state 2, as shown on Figure \ref{Fig3-beta2} (top panels). In the top panels of Figure \ref{Fig3-beta2}, the electron $\beta_e$ computed at the distance of $35.6 R\odot$ is plotted against the proton speed for calm solar wind intervals (top left panel) and for the rest of the data-set (top right panel), in red (magenta) for state 1 and in blue (cyan) for state 2. In each of the two top panels, the difference between the two plasma states is obvious. In both states, the solar wind can be considered as a "slow" solar wind, even if the plasma reaches much higher speed values (almost twice higher) in state 2. In state 1, the plasma gas pressure dominates the magnetic pressure. In state 2, on the contrary, the magnetic pressure dominates over the gas pressure.

Figures \ref{Fig2-1}, \ref{Fig2-2} and \ref{Fig3-beta1}, reveal that the parameter with the greatest change between the two states is the plasma density. Figure \ref{Fig3-beta2} shows the electron density scaled to $35.6 R\odot$ (middle panels), with the error bars provided by the plasma QTN spectrum measured by RFS/FIELDS \citep{moncuquet_first_2020}, which are used to compute error bars on the $\beta_e$, as the other quantities are not provided with errors. On the left middle panel, data from calm solar wind are plotted in red for state 1, and in blue for state 2. Those calm solar wind intervals are representative of a "background" solar wind, plotted against the proton speed, which is representative of the plasma bulk speed. The electron density is higher across the speed range in state 1 than in the second one. In state 1 (in red), the faster the plasma is, the higher is the electron density. In state 2 (in blue),  an opposite behaviour is observed: the electron density decreases when the plasma speed increases. This statement is supported by linear least square fit, which provides a positive slope of +0.31 in state 1 and a negative one of -0.23 in state 2. The anti-correlation is strongest on speeds between 260 and 370 km/s. Beyond 350 km/s, the linear least squares on state 2 data shows that the density is constant as the speed increases. The unscaled electron density
 shows the same trends. \\

Magnetic field is stronger in state 2. There is no particular relationship between the magnetic field and the plasma speed, as can be seen on Figure \ref{Fig3-beta2}, bottom panels.  Figure \ref{Fig3-beta2} (bottom panels) shows magnetic field intensity scaled to $35.6 R\odot$. There is a clear difference between the two plasma states when regarding the calm solar wind intervals (left bottom panel), essentially due to the fact that the magnetic field is stronger in the plasma state encountered after April 3, 2019, at 07:55 UT. The rest of the data-set (right bottom panel) reveals that the presence of magnetic field fluctuations (including magnetic switchbacks) make the two states less distinct between each other from the point of view of the magnetic field intensity. Variability of the magnetic field is much higher in state 1 than in state 2. \\

\subsection{Temperatures}

Electron core temperature is relatively constant over all the data-set under consideration in this study (see third panel from top of Figure \ref{Fig2-1}), i.e. from March 31, 2019, 00:00 UT to April 5, 2019, 23:59 UT. Electron core temperature is thus not responsible for the two different plasma states. However, small differences in terms of temperatures are still noticeable from one state of plasma to the other.

\begin{figure*}[!h]
	\includegraphics[width=\linewidth]{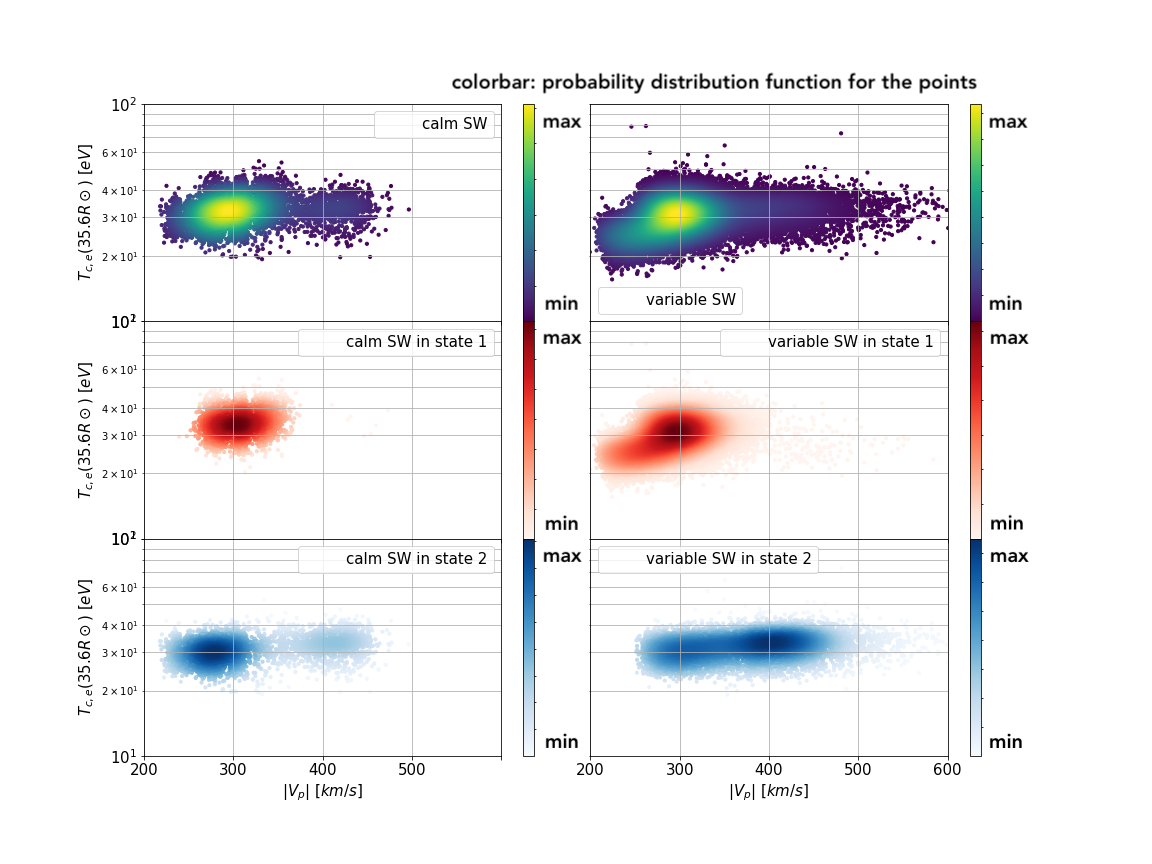}
	\caption{Core electron temperature $T_{c,e}$ (in eV) scaled to $35.6\ R_\odot$, provided by the plasma QTN spectrum measured by RFS/FIELDS, against plasma speed (km/s), provided by SWEAP/SPC. Left column: data for calm solar wind intervals only, for the entire data-set (top), in state 1 (middle, red-to-black colours) in state 2 (bottom, blue-to-black colours). Right panel: same organisation but for rest of the dataset. Colours indicate the probability distribution function for the points.	}
	\label{Fig3-temp1}
\end{figure*}

When considering Temperature-speed diagrams for the core electron temperature $T_{c,e}$ and proton speed $V_{p}$ (used as a proxy of plasma speed) on Figure \ref{Fig3-temp1}, it is interesting to compare the global T-V diagrams (for all data points, top panels), in state 1 on the middle panel and in state 2 on the bottom panel, and note that the "gun-shaped" distribution is actually made up by the sum of two distributions, corresponding respectively to the data in state 1 (middle panels, in red-to-black colours) and data in state 2 (bottom panels, in blue-to-black colours). T-V diagrams considered separately present slightly different slopes from one state of plasma to the other (on calm solar wind intervals, the slope is $0.04\pm0.08$ in state 1 and $0.02\pm0.07$ in state 2), resulting in the "gun-shaped" global distribution. Moreover, T-V diagrams were plotted (but not displayed) for both core electron temperature $T_{c,e}$ and proton temperature $T_{p}$, showing the expected difference between electron and proton temperatures: when electron temperature is relatively constant with plasma speed, the proton temperature actually increases with the plasma speed. The speed at which protons become hotter than electrons is around 325 km/s in state 1, and around 380 km/s in state 2. The slope of the T-V distribution for proton temperature seems to be steeper in state 1. Although, the major change in electron density from state 1 to state 2 could affect the absolute temperature measurements. A complete study of the T-V diagrams is out of the scope of this paper, but should be investigated in order to establish how the different slopes could be relevant of different heating/propagation mechanisms. \\

\subsection{Electron flux vs magnetic pressure }

\begin{figure*}[!h]
	\includegraphics[width=\linewidth]{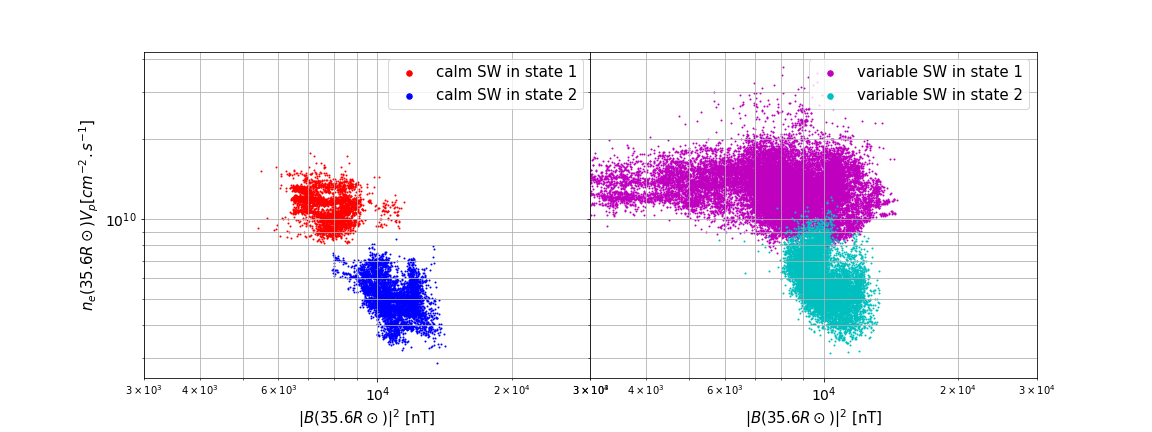}
	\caption{Electron 'flux' $n_{e}(35.6 R\odot)V_p$ (in $cm^{-2}.s^{-1}$), provided by the plasma QTN spectrum measured by RFS/FIELDS and SWEAP/SPC, against the square magnetic field intensity (in nT), provided by FIELDS. Left panel: data for calm solar wind interval (red for state 1, blue for state 2). Right panel: data for variable solar wind intervals (magenta for state 1, cyan for state 2).}
	\label{Fig3-emf-vs-magpres1}
\end{figure*}

The two different plasma states show clearly different fluxes, approximated here as $n_{e}(35.6 R\odot)V_p$ (in $cm^{-2}.s^{-1}$), as shown in Figure \ref{Fig3-emf-vs-magpres1}. The higher speed in state 2 does not compensate the sudden drop in plasma density, thus the flux in state 2 is lower. This is true for both calm solar wind intervals and for variable solar wind intervals with numerous switchbacks. Plotted against the square of the magnetic field intensity (representative of magnetic pressure) two plasma states appear distinct both in terms of the gas and magnetic field properties. There is a slow solar wind characterised by low magnetic pressure and high flux and another slow wind with low flux and high magnetic pressure (see figure \ref{Fig3-emf-vs-magpres1}). 

\newpage

\section{To the source of the wind}
\label{section:m&m-connectivity}

\subsection{Detailed analysis of PSP's connectivity}

\begin{figure*}[!t]
		\includegraphics[width=0.8\linewidth]{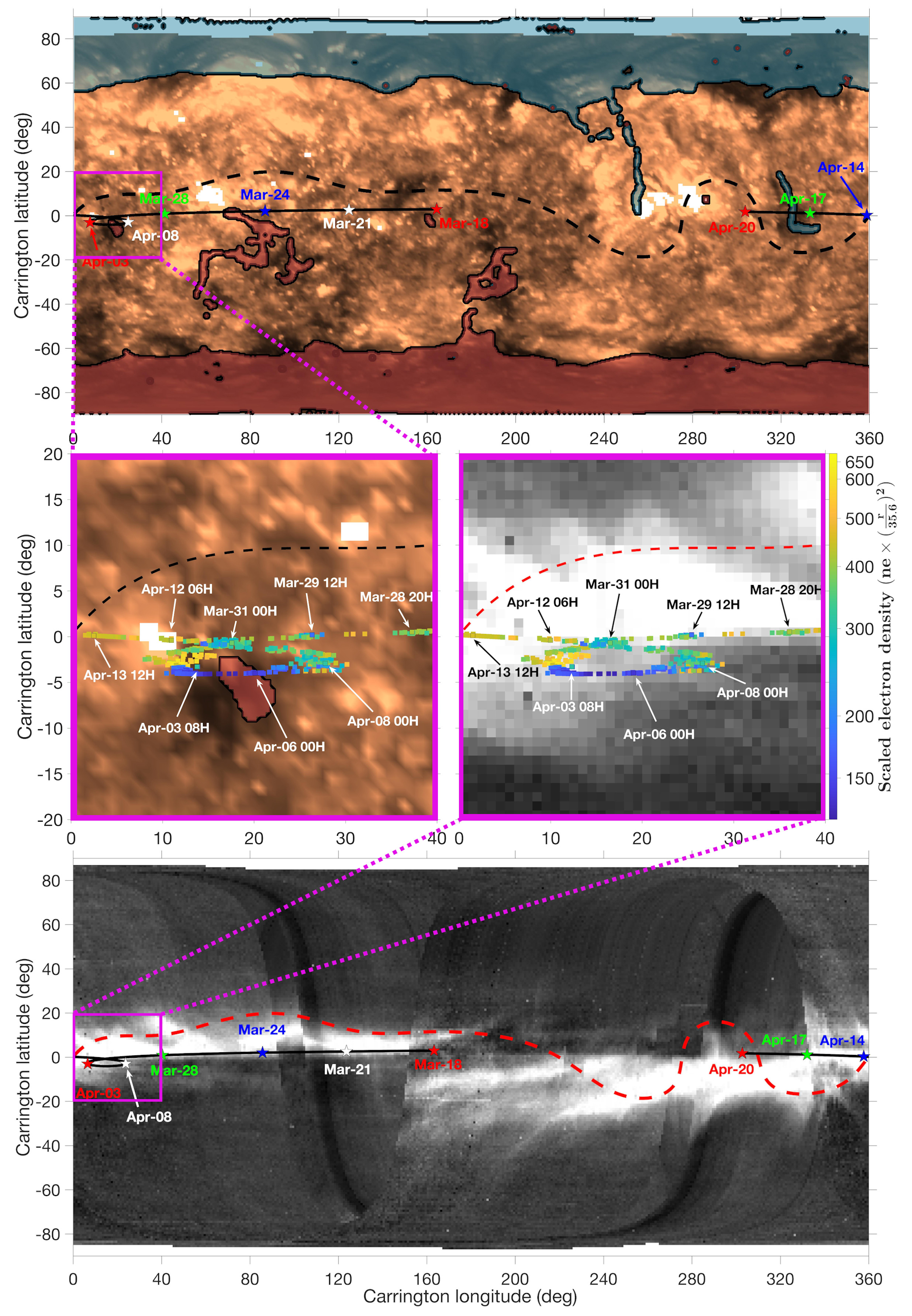}
		\caption{Top panel: Carrington map from SDO-AIA EUV observations, on March 31, 2019, at 00:00 UT and at 193 \AA. Bottom panel: Carrington map from SOHO-LASCO C3 white-light observations, on April 3, 2019, at 12:00 UT and at a height of 5 $R\odot$. Both Carrington maps, although they are synoptic maps over the full solar rotation, are continuously updated as new images are available. The time label associated to the Carrington maps therefore corresponds to the date (and time) of the last update. The PSP trajectory (dark solid-line) is projected at the height of the Carrington maps (and at 2.1 $R\odot$ for the AIA EUV map), using the Parker Spiral \citep{parker_dynamics_1958} and the solar wind speed measured in situ at PSP. Along the PSP projected trajectory, some PSP projected positions for several dates and times are depicted with colored stars and arrows. On the zoom-in views on the middle panels, the plasma electron density $n_{e}(35.6 R\odot)$ (in $cm^{-3}$) (FIELDS/RFS(QTN)) is plotted along the PSP projected trajectory with a logarithmic color scale. The red and dark dashed-lines represent the neutral line from the PFSS model (described in section \ref{section:m&m-connectivity}). Open field regions of positive (negative) polarity from the PFSS model are represented at the photosphere by a semi-transparent blue (red) layer.
		}
		\label{Fig3-comparison_WLimages}
	\end{figure*}

In the previous sub-sections, several macroscopic properties of the solar wind were compared before and after April 3, 2019, at 07:55 UT (plasma density, speed, temperature, flux, beta, and magnetic field intensity). Those differences suggest that Parker Solar Probe (PSP) crossed, successively, two different states of slow solar wind, for at least three days before April  3, 2019, at 07:55 UT (state 1) and three days after (state 2). As mentioned in Section \ref{section:introduction}, this date of April 3, 7:55   UT was first investigated in a study of white-light images of the corona taken by SOHO and STEREO-A. It revealed that PSP escaped streamer flows to sample solar wind likely released from deeper inside a coronal hole \citep{rouillard_relating_2020}. The latter study did not investigate the presence of such a coronal hole that we address here in more detail. A more precised study of PSP connectivity against white-light Carrington maps was therefore pursued in the present study in order to characterise potential sources of the slow wind and is presented in the top panels of figure \ref{Fig3-comparison_WLimages}. 

We assumed that Parker Solar Probe was connected to the upper corona by a Parker spiral defined by the speed of the solar wind measured in situ. We then connected the upper corona to the photosphere by using a Potential Field Source Surface (PFSS) extrapolation of the photospheric magnetograms \citep{wang_potential_1992}. There is considerable uncertainty associated with determining the exact configuration of the solar magnetic field in order to determine how a spacecraft connects magnetically to the different regions of the solar atmosphere \citep{rouillard_models_2020}. This is related to the inherent assumptions of the model such as a spherical shape of the source surface or the assumption of a current-free corona. Any model of the solar atmosphere is also dependent on the quality of solar magnetograms and the choice of free parameters. In the PFSS model this is the height of the source surface \citep[see e.g.][]{badman_magnetic_2020, panasenco_exploring_2020}. White-light observations from the SOHO-LASCO C2 and C3 instruments can be exploited to extract the location of the streamer belt and to compare this location with that of the polarity inversion line obtained with each PFSS realisation. We developed a new technique to segregate between PFSS realisations obtained for different choices of solar magnetograms and source surface heights based on how well the derived neutral line position matched the location of the streamer belt. The streamer belt is reduced to a single curve defined by the locus of maximum brightness, hereafter called Streamer Maximum Brightness (SMB) line. For most longitudes the SMB  lies at the centre of the streamer belt. The angular distance between the SMB line and the neutral line from the PFSS model can be measured. We then consider that a PFSS realisation fits best the white-light observations when the distance between the SMB and the neutral line averaged over all relevant longitudes is minimal. This comparison is prioritised in the range of Carrington longitudes magnetically connected to PSP via the Parker spiral. The height of the source surface was further adjusted until the small equatorial coronal hole observed in EUV was reproduced by the PFSS model (data from SDO-AIA, 193\AA). In this iterative process, a compromise had to be found in order to keep a PFSS neutral line locally consistent with the white-light observations. The best-fit PFSS model was obtained for a GONG-ADAPT magnetogram (realization n$^\circ$2) of March 24, 2019, at 00:00 UT and a source height of 2.1 $R\odot$. The GONG magnetogram was processed by the Air Force Data Assimilative Photospheric Flux Transport model (ADAPT) which simulates the motion of photospheric magnetic fields \citep[see][]{arge_improved_2003}. The magnetic connectivity from the source surface (at 2.1 $R\odot$) was then extended using the Parker spiral model \citep{parker_dynamics_1958} for the solar wind speed measured in situ at PSP. \\

From these maps, it is clear that PSP left the boundary of the streamer around April 3, 2019, before reentering it at around April 6. Our modelling of PSP's connectivity around April 3, 2019 reveals that the spacecraft has intersected magnetic field lines rooted deep inside an equatorial coronal hole, clearly visible (as a dark area) on Extreme UV (EUV) light maps (middle left panel in Figure \ref{Fig3-comparison_WLimages}) during this period. This small coronal hole remains visible in the EUV map during the entire time interval considered here, though varying slightly over time. 

\subsection{Numerical simulations of the slow solar winds}

In order to link the magnetic field configuration obtained from the optimised PFSS model and the properties of plasma states 1 and 2, we use the magnetohydrodynamic "single fluid" MULTI-VP code \citep{pinto_multiple_2017}, that computes full solar wind solutions (surface to high corona) taking into account heating and cooling processes in the corona. In MULTI-VP, as explained in \cite{pinto_multiple_2017} and \cite{griton_coronal_2020}, the mechanical heating directly depends on the photospheric magnetic field. We associate the mechanical heating flux $F_{\rm h}$ (or, equivalently, the heating rate $Q_h = -\nabla\cdot F_h$) to a function that represents the effects of the coronal heating processes:
\begin{equation}
  \label{eq_flux_global}
  F_{\rm h} = F_{\rm B0} \left(\frac{A_0}{A\left(s\right)}\right)h\left(s\right),
\end{equation}
with $A\left(s\right)$ being the flux tube's cross sectional area at a given curvilinear coordinate $s$, $A_0$ the flux tube's cross sectional area at the solar surface, and the coefficient $F_{\rm B0}$ proportional to the basal magnetic field amplitude $\left|B_0\right|$ (with, for basal field amplitudes consistent with typical Wilcox Solar Observatory magnetograms, $F_{\rm B0}=8 \times 10^5 \left|B_0\right|\ \mathrm{erg\ cm^{-2}\ s^{-1}}$). The coronal heating profile $h\left(s\right)$ is
\begin{equation}
  \label{eq_fluxp}
  h\left(s\right)=\exp\left[-\frac{s-R_\odot}{H_{\rm f}}\right],
\end{equation}
which takes into account a damping scale-height $H_{\rm f}$, which is anti-correlated with the superradial expansion ratio in the low corona, as in \citet{pinto_multiple_2017}.

\begin{figure*}[!h]
	\includegraphics[width=\linewidth]{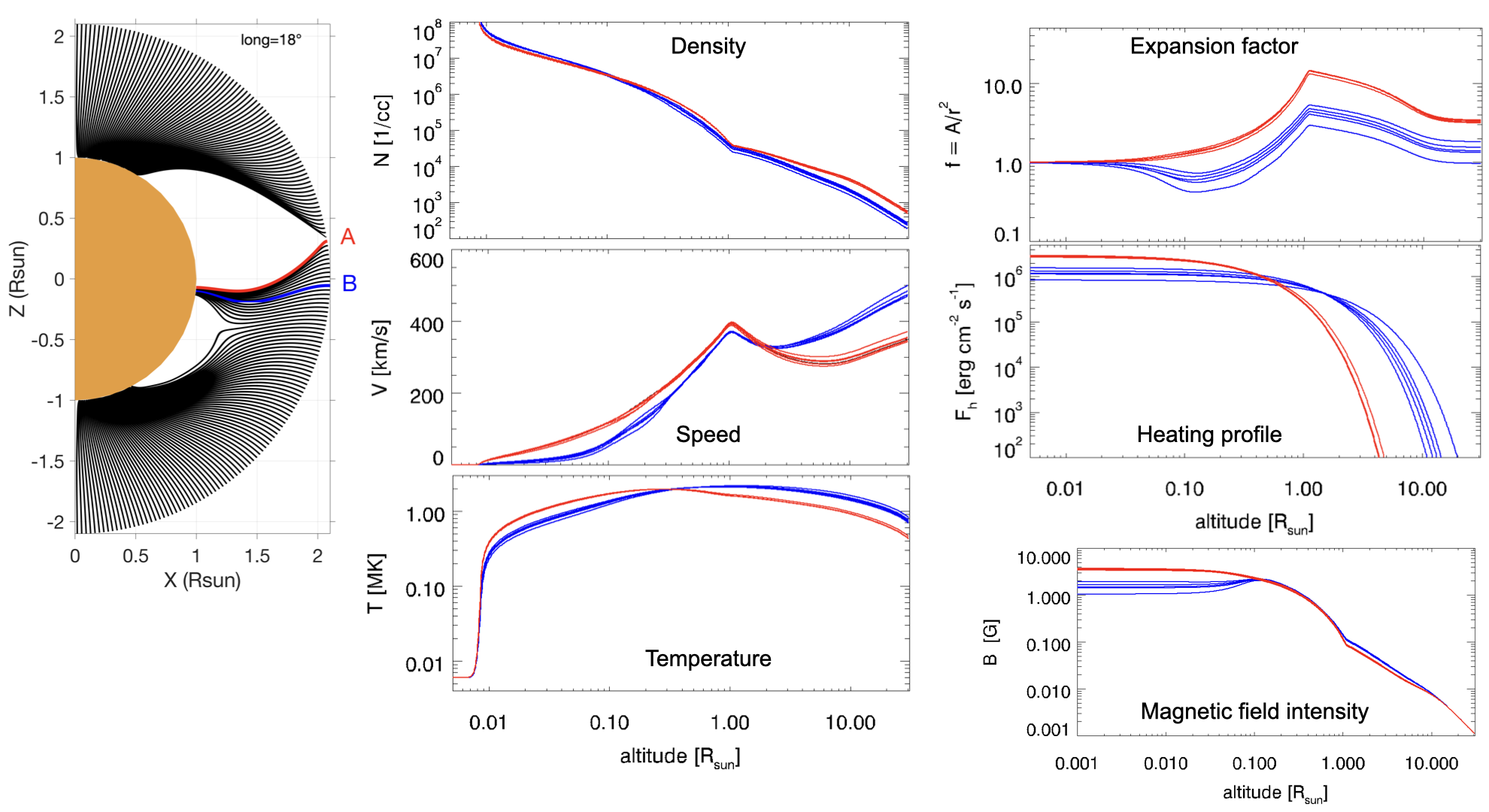}
	\caption{Left: Samples of magnetic field lines from the optimised PFSS magnetic field model with the source surface at 2.1 $R\odot$, in the plane cut through $18\degree$ longitude. PSP certainly crossed magnetic field lines connected to field lines A (state 1) and B (state 2). Middle: plasma density (top), speed (middle) and temperature profiles computed by the MHD code along 5 flux tubes around field line A (in red) and around field line B (in blue). Right: expansion factor (top), heating profile (middle) and magnetic field intensity profiles computed by the MHD code along 5 flux tubes around field line A (in red) and around field line B (in blue). }
	\label{Fig4-context}
\end{figure*}

Thus MULTI-VP computes the profiles of contiguous one dimensional solar wind streams from the surface of the Sun up to about $30\ R_{\odot}$. The individual streams are guided along individual magnetic flux tubes, which are obtained from the optimised PFSS model: we manually identify the magnetic field tube A as representing the edge of streamer tubes connected to PSP from March 31 to April 3 2019, and the magnetic field tube B  for the egde of the equatorial coronal hole. A projection of the field line trajectories into a meridional plane at $\Phi=18 \degree$ Carrington longitude is displayed on the left panel of figure \ref{Fig4-context}, where magnetic field lines A and B are highlighted in red and blue.

The middle and right panels figure \ref{Fig4-context} present the results of the MULTI-VP simulations along five flux tubes around tube A (red profiles) and five flux tubes around tube B (blue profiles), separated by 1 degree of spatial resolution at the source surface, at 2.1 $R\odot$ from the Sun center. The simulations reproduce many properties measured in situ (and usually produced by MULTI-VP simulations \citep{pinto_multiple_2017}): the solar wind along tube A, representative of plasma state 1, is indeed slower and denser compared with the solar wind along tube B, representative of plasma state 2 and emerging from the equatorial coronal hole. On the logarithmic scale, a kink lies at about $h=1R\odot$ (i.e, at a radial distance of 2.0$R\odot$ from the center of the Sun), still slightly below the source-surface radius at 2.1$R\odot$. This difference of almost 0.1 $R\odot$ is correct and occurs because the PFSS extrapolation produces sometimes very sharp fieldline bends just before they connect to the radial field above. The numerical code requires that all quantities are spatially resolved and are smooth enough (i.e, 3rd order differentiable). For safety, a thin interface zone is introduced in between the two coronal regions, on which the maximum gradients of the fieldline profiles are constrained. That interface region is 0.1 $R\odot$ wide, and only acts effectively on rather extreme field line geometries. The overall representation of such field lines remains faithful to that of the underlying extrapolation.
The presence of a pseudo-streamer (visible on the cut on the left panel of figure \ref{Fig4-context}) bends both flux tubes in such a way that the expansion of both tubes A and B exert a significant deviation from the radial direction around an altitude of $0.5\ R_{\odot}$, with an effect on the number density and speed profiles at this altitude. Temperature profiles provide an explanation for the differences between the two states of solar wind. As we said, the heating depends on the photospheric magnetic field but also on the geometry of the magnetic flux tube. Heating more concentrated lower down (due to the geometry) makes the lowest and densest layers expand (that is, increase scale-height, or  "evaporate"). There are then more particles per "unit energy" although the thermal driving is lower. As a result, along streamer-like flux tubes, density, temperatures and speeds are lower higher in the corona for a similar energy flux.

\section{Discussion}
\label{section:discussion}

This study focused on two states of plasma encountered by Parker Solar Probe (PSP) from March 31, 2019, 00:00 UT to April 3, 07:55 UT (state 1) and from April 3, 07:55 UT to April 5, 2019, 23:59 UT (state 2). In order to trace the origins of these solar winds, we carried out a combined analysis of in situ measurements, remote-sensing observations with a modeling effort to select the most appropriate 3-D magnetic field reconstructions, and numerical simulations of the wind between the solar surface and 30$R\odot$.  Our results suggest that the two states present very different properties which are characteristic of two different magnetic field configurations in the vicinity of the Sun (below 2.1 $R\odot$): state 1 originates from open magnetic field lines rooted on the boundary of the equatorial coronal hole and that expand significantly over streamer loops before reaching the upper corona. State 2 emerges along open fields that root deeper inside an equatorial coronal hole but do not expand as much. 

A meridional cut of the magnetic field reconstructed from our optimised PFSS model, described in section \ref{section:m&m-connectivity}, is shown for heights below $2.1R\odot$ in figure \ref{Fig4-context} and inside the range of Carrigton longitudes sampled by PSP. State 1, identified as emerging from the vicinity of the main streamer (typically magnetic field line A, in red), presents a denser plasma, embedded in a lower magnetic field than state 2, identified as emerging from an equatorial coronal hole (typically magnetic field line B, in blue on figure \ref{Fig4-context}). This observation is coherent with past observations of plasma coming from the tip of streamers and pseudo streamers compared with the solar wind emerging from coronal holes \citep{cranmer_origins_2017, sanchezdiaz_very_2016,withbroe_temperature_1988}. If not due to calibration issues, temperature-speed (T-V) relationships revealed that the slope of the T-V relationship is different from state 1 to state 2, and that the speed at which the temperature of protons exceeds the temperature of electrons (which is of 500 km/s at 1 AU \citep{newbury_electron_1998}) is faster in the coronal-hole-like plasma (state 2) than in the streamer-like plasma (state 1). From the single fluid simulations, it seems that plasma state 2 (i.e. originating from the coronal hole) is hotter and faster than plasma state 1 at a given altitude above the source surface. As from PSP data, electron temperature is not changing from plasma state 1 to plasma state 2, it can be deduced that the change in temperature observed in the simulations has to be supported by protons in the real solar wind. Thus, the speed at which proton temperature exceeds that of the electrons should be systematically higher in state 2 than in state 1. This suggests that the T-V relation measured in the slow winds bears some dependence on the source region, in addition to the radial evolution already stated by \cite{elliott_temporal_2012} and revisited through PSP data by \cite{maksimovic_anticorrelation_2020}. As explained by \cite{demoulin_why_2009}, the T-V relationship in plasma flowing along open magnetic field lines can provide information on the acceleration and heating processes near the Sun, which could suggest that states 1 and 2 are built through different heating processes very low in the solar corona. In the numerical simulations (see profiles on Figure \ref{Fig4-context}), this difference of temperature is clear above $2.1R\odot$. As in red profiles the magnetic field is more intense from the photosphere to $0.1R\odot$, and as the expansion factor is higher at almost all altitudes between $0R\odot$ and $1R\odot$, the heating flux, defined by equation [\ref{eq_flux_global}], remains elevated in red profiles below the altitude of $1.1R\odot$. This explains why the temperature is higher along red profiles than along blue profiles below $1.1R\odot$. However, above $2.1R\odot$ from the center of the Sun, the heating flux drops much faster (i.e. at lower altitude) along red profiles compared to blue profiles. Moreover, the stronger heating at the coronal base has increased the rate of chromospheric evaporation which means that more matter travels along red profiles in the upper corona (as revealed by the density profiles).  Beyond the altitude of $1.1R\odot$, the heating flux is low due to the high expansion factor which reduces the available energy per particle, resulting in lower temperatures along red profiles than along blue profiles above the Potential Field Source Surface. A quick analysis of the proxy of suprathermal electrons' temperature revealed that state 2 may contain cooler suprathermal electrons than state 1, which would be coherent with the analysis from \cite{bercic_coronal_2020} conducted on PSP first perihelion, when the spacecraft was briefly connected to an equatorial coronal hole \citep{bale_highly_2019}.

The choice of magnetogram and source surface radius that allows the PFSS model to best fit the white light images and EUV observations was essential to identify the presence of the equatorial coronal hole. This method is limited when it comes to reproduce the exact size of the coronal hole and compare PSP data with the magnetic field at source surface with a time resolution lower than half a day. However, it is important to include this equatorial coronal hole in the reconstruction of the solar magnetic field used to run the MULTI-VP simulations. The magnetic field at the photosphere strongly differs from one type of field line (i.e. streamer-like at the boundary of the coronal hole) to another (i.e. deeper inside the equatorial coronal hole) even if their footpoints are close on the photosphere. This fact is supported by the red (representative of state 1 plasma) and blue (state 2) magnetic field profiles on figure \ref{Fig4-context}. We note that there is no "reconnection" in the PFSS model, and the differences in the coronal magnetic fields are therefore independent of any reconnection process. \\

Calm solar wind intervals, with no magnetic switchbacks, were clearly identified in both types of solar winds revealing what appears to be an undisturbed "background" solar wind that emerges from both streamer flows and from deeper inside coronal holes. Radial evolution was removed at first order using known heliocentric variations of magnetic field, plasma density and temperatures in the solar wind. Focusing on the background solar wind and removing first order radial distance is essential to study the link between plasma properties measured at PSP, around $35.6R\odot$, and the configuration in the vicinity of the Sun that could be responsible for the formation of different plasma states.

Modelling of the formation of the solar wind \citep{pinto_multiple_2017, griton_coronal_2020} shows that different plasma density and speed regimes can be produced along flux tubes with different geometries, when the heating in the corona is made dependent on the local magnetic field intensity (or expansion profiles). On a flux tube close to the streamer (flux tube A on figure \ref{Fig4-context}), the heating is essentially deposited very low, near the surface, which produces a dense and slow solar wind. On the contrary, the heating is more evenly distributed along the flow tube B producing a less dense but faster solar wind than along tube A. When looking carefully at the second panel of figure \ref{Fig2-2}, one can note that the speed increases gradually, over two days, and reaches around 450 km/s on April 5. This evolution could be the consequence of PSP getting closer to the center of the equatorial coronal hole. The frontier between the two plasma state here is very well-defined and sharp compared to any boundary between streamer-like plasma and equatorial coronal hole plasma encountered on the first perihelion of PSP \citep{bale_highly_2019}. We expect the topology of the equatorial coronal hole encountered on April 3rd to be more compact (i.e. smaller  in diameter and more intense in magnetic field), such that PSP crossed it deeper and faster than the first encounter equatorial coronal hole, resulting in the sharp boundary we observe on density and speed profiles.

The fact that calm wind intervals are identified inside streamer flows close to the source is supporting the idea that dynamic processes near the tip of streamers are not necessary to produce the dense and very slow solar wind measured near streamers \citep{sanchezdiaz_very_2016}. Nevertheless the solar wind measured before April 3, 7:55   UT (state 1) remains overall significantly more variable than the solar wind originating from deeper inside the coronal hole (state 2). The more variable part of the streamer flow is likely due to continual magnetic reconnection occurring between the open field lines and the loops of the streamer. \\

The slow solar wind emerging from deeper inside the equatorial coronal hole (state 2) has more frequent periods of very quiet wind with no switchbacks. The switchbacks in this region tend also to be less pronounced as measured by the duration of the inversion in magnetic field direction. Perhaps magnetic reconnection in this region could be sustained by continual magnetic reconnection with much smaller closed loops that would typically occur for low-latitude coronal holes, allowing the coronal holes to rotate quasi-rigidly with their extensions higher in the solar corona while their foot-points are moving along with the photospheric magnetic field \citep{nash_mechanisms_1988, fisk_motion_1996, fisk_coronal_1999}.

An alternative viewpoint was proposed recently for the origin of switchbacks and is related to the radial evolution of turbulent flows \citep{squire_-situ_2020}. In the latter scenario, the intense magnetic switchbacks measured in the streamer flow could be formed in situ as a result of the significant expansion of magnetic field lines that form the helmet streamer. The weaker switchbacks measured in the coronal hole flow could be related to the smaller expansion factors of these associated coronal magnetic field.

\section{Conclusion}
\label{section:conclusion}
In summary, our work investigated two plasma states of slow solar wind measured by Parker Solar Probe from March 31, 2019, 00:00 UT to April 3, 2019, 07:55 UT (state 1: streamer flow) and from April 3, 2019, 07:55 UT to April 5, 2019, 23:59 UT (state 2: coronal hole). The major findings from this work are:
\begin{enumerate}
	\item the solar wind from a streamer has higher flux and thermal pressure but lower magnetic field pressure than the solar wind from deep inside the equatorial coronal hole. This results in different plasma $\beta$, with the highest plasma beta measured in the streamer flows,
	\item the streamer flow is more perturbed by switchbacks than the coronal hole flow, however we have found time intervals lasting several hours during which no switchbacks were measured. These very quiet solar wind intervals suggest that slow solar wind can form along streamer stalks and deeper inside isolated coronal holes without significant transient activity near its source,
	\item detailed modelling allowed us to establish a link between the two wind states and two magnetic properties of their coronal sources. We associate the different slow wind states in part to the coronal height at which energy is injected in the corona. The dense and slow streamer flow is likely to result from coronal heating confined mainly to the very low corona driving a high flux, as suggested by numerical simulations. This wind does not accelerate significantly due to the highly expanding magnetic flux tube higher in the corona. This results in a solar wind with high flux and thermal pressure but low magnetic field pressure. In contrast the slow wind that originates along magnetic field lines rooted closer to the center of the equatorial coronal hole do not expand as much. Heating occurs along a more extended section of the flux tube and results in a small but longer lasting acceleration. 
\end{enumerate}

In this study we did not investigate the origin of the switchbacks that appear to have different properties in the two slow wind states. We consider that calm solar wind intervals are representative of the "background" solar wind that is emitted from the Sun, whose properties are depending only on the configuration of the magnetic field and not on sporadic events induced by magnetic reconnection. In other words, switchbacks could constitute a variability that develops from a background solar wind, which is revealed during the calm solar wind intervals. We note that equatorial coronal holes appear and disappear during the rising phase of a solar cycle as non-axisymmetric magnetic field emerges at low latitudes \citep{Wang2010}. Therefore we expect the boundary of equatorial coronal holes to be dynamic and perturb the background solar wind emerging from these regions. This could induce the strong variability measured in state 1. Such investigation is kept for a future publication. More definitive answers to the origins of the slow winds will likely come with the comprehensive composition measurements taken by Solar Orbiter since composition is invariant during the solar wind's propagation.\\

\begin{acknowledgements}

The IRAP team acknowledges support from the French space agency (Centre National des Etudes Spatiales; CNES; https://cnes.fr/fr) that funds activity of the space weather team in Toulouse (Solar-Terrestrial Observations and Modelling Service; STORMS; http://storms-service.irap.omp.eu/). The work of L. Griton, A.P.R., and N.P. was funded by the ERC SLOW\_SOURCE project (SLOW\_SOURCE—DLV-819189). We acknowledge the NASA Parker Solar Probe Mission, the
FIELDS team led S. D. Bale, and the SWEAP team led by J. Kasper for use of
data. The FIELDS and SWEAP experiments on the Parker Solar Probe spacecraft
were designed and developed under NASA contract NNN06AA01C.
\end{acknowledgements}

\bibliographystyle{apj} 
\bibliography{source-dependent-properties} 

\end{document}